\documentclass[7pt,a4paper,twocolumn]{article}
\usepackage[utf8]{inputenc}
\usepackage[T1]{fontenc}
\usepackage{mathptmx}
\usepackage[pdftex]{graphicx}
\usepackage[pdftex,linkcolor=black,pdfborder={0 0 0}]{hyperref}
\usepackage{calc}
\usepackage{enumitem}
\usepackage{tikz}

\usepackage[a4paper, lmargin=0.08\paperwidth, rmargin=0.08\paperwidth, tmargin=0.08\paperheight, bmargin=0.08\paperheight]{geometry}

\usepackage[all]{nowidow}
\usepackage[protrusion=true,expansion=true]{microtype}

\usepackage{amsmath}
\usepackage{amsthm}

\usepackage{amsfonts}
\usepackage{complexity}
\usepackage{xurl}

\usepackage{xcolor}

\DeclareRobustCommand{\doi}[1]{%
  \href{https://doi.org/#1}{\colorbox{gray!20}{\texttt{DOI}}}%
}

\usepackage{enumitem}
\usepackage{algpseudocode}

\usepackage{titlesec}
\titlespacing*{\section}{0pt}{1pt}{1pt}
\titlespacing*{\subsection}{0pt}{1pt}{1pt}
\titlespacing*{\paragraph}{0pt}{1pt}{1pt}

% ...
\usepackage{lipsum}

\author{Kühn, Kolja}
\title{Computing Treedepth Obstructions}
\makeatletter

\hypersetup{
    pdfsubject = {Description of your ORCS.},
    pdftitle = {\@title},
    pdfauthor = {\@author}
}

\usepackage{fancyhdr}
\newcommand{\threeObstruction}{
    \begin{tikzpicture}[every node/.style={circle,draw,fill,minimum size=3mm,inner sep=0pt}]
        \node (a1) at (0,1) {};
        \node (a2) at (1,1) {};
        \node (a3) at (2,1) {};
        \node (a4) at (3,1) {};
        
        \node (b1) at (0,0) {};
        \node (b2) at (1,0) {};
        \node (b3) at (2,0) {};
        \node (b4) at (3,0) {};
        
        \draw (a1)--(a2)--(a3);
        \draw (b2)--(b3)--(b4);
        
        \draw (a1)--(b1);
        \draw (a2)--(b2);
        \draw (a3)--(b3);
        \draw (a4)--(b4);
    \end{tikzpicture}
}

\begin{document}
    \pagenumbering{gobble}
    An elimination forest $F$ for a graph $G = (V,E)$ is a rooted forest on the vertex set $V$ such that for every edge $uv \in E$, $u$ is an ancestor or descendant of $v$ in $F$.
    The treedepth $\text{td}(G)$ of a graph $G$ is defined as the minimum height of an elimination forest for $G$.
    Treedepth is monotone under taking minors, thus, the class $\mathcal{G}_k$ of graphs with treedepth at most $k$ is a minor-closed class.
    From the Graph-Minor-Theorem \cite{robertson2004graph} it follows that there is a finite set $\text{Obs}_\leq(\mathcal{G}_k)$ of forbidden minors that characterize $\mathcal{G}_k$.
    Dvo{\v{r}}{\'a}k et al. \cite{dvovrak2012forbidden} computed $\text{Obs}_\leq(\mathcal{G}_3)$, and furthermore showed that for every $k$ there exist obstructions for $\mathcal{G}_k$ that contain $2^k$ vertices.
    They conjecture that this is an upper bound, i.e., that every graph $G \in \text{Obs}_\leq(\mathcal{G}_k)$ has at most $2^k$ vertices, as well as identical bounds for the set of forbidden subgraphs $\text{Obs}_\subseteq(\mathcal{G}_k)$ and forbidden induced subgraphs $\text{Obs}_\sqsubseteq(\mathcal{G}_k)$.

    For a graph class $\mathcal{G}$, we define $\mathcal{G}^{(n)} := \{G \in \mathcal{G} \colon |V(G)| = n\}$ and $\mathcal{G}^{(\leq n)} := \{G \in \mathcal{G} \colon |V(G)| \leq n\}$.
    We compute for each $n \leq 16 = 2^4$, the set of $n$-vertex obstructions for treedepth at most $4$, $\text{Obs}^{(n)}_R(\mathcal{G}_4)$ with $R \in \{\leq, \subseteq, \sqsubseteq\}$.
    If the conjecture holds, these are the complete sets for treedepth at most $4$.

    \subsection*{The Algorithm}
    We sketch an algorithm that for $k, n \in \mathbb{N}$ computes $\text{Obs}^{(n)}_R(\mathcal{G}_k)$ with $R \in \{\leq, \subseteq, \sqsubseteq\}$.
    We use the fact that $\text{Obs}_\sqsubseteq^{(n)}(\mathcal{G}_k) \supseteq \text{Obs}_\subseteq^{(n)}(\mathcal{G}_k) \supseteq \text{Obs}_\leq^{(n)}(\mathcal{G}_k)$ to obtain the latter two sets by filtering the first set.
    \setlist{nolistsep}
    \begin{enumerate}[itemsep=0em]
        \item Recursively compute $\mathcal{G}_k^{(i)}$ for $1 \leq i \leq n - 1$.

        \item Compute $\text{Obs}_\sqsubseteq^{(n)}(\mathcal{G}_k)$ as the set of graphs $G$ satisfying: $\text{td}(G) = k + 1$ and $\forall v\in V(G) \colon G - v \in \mathcal{G}_{k}^{(n - 1)}$.
        
        \item Compute $\text{Obs}_\subseteq^{(n)}(\mathcal{G}_k)$ as the set of graphs that are subgraph-minimal in $\text{Obs}_\sqsubseteq^{(n)}(\mathcal{G}_k)$.
        
        \item Compute $\text{Obs}_\leq^{(n)}(\mathcal{G}_k)$ as the subset of graphs from $\text{Obs}_\subseteq^{(n)}(\mathcal{G}_k)$ that are minor-minimal in $\text{Obs}_\subseteq^{(n)}(\mathcal{G}_k) \cup \text{Obs}_\sqsubseteq^{(n - 1)}(\mathcal{G}_k)$.
    \end{enumerate}

    In \textbf{Step 1} we initialize $\mathcal{G}_k^{(1)} := \{K_1\}$.
    Given $\mathcal{G}_{k}^{(i - 1)}$, we compute $\mathcal{G}_k^{(i)}$ as the union over all single-vertex extensions of graphs in $\mathcal{G}_{k}^{(i - 1)}$:
    For each $G = (V,E) \in \mathcal{G}_{k}^{(i - 1)}$ and every set of vertices $A \subseteq V$, we define the graph $G_{A} := (V \cup \{v\}, E \cup \{uv \mid u \in A\})$.
    We may assume without loss of generality that $v$ is a vertex of minimum degree in $G_{A}$, thereby greatly reducing the number of vertex sets $A$ that need to be considered.
    We then test whether $\text{td}(G_{A}) \leq k$, compute a canonical representation $G_{A}^*$ of $G_{A}$, and store $G_{A}^*$ in a hash-set, allowing us to eliminate duplicates.

    \textbf{Step 2} is implemented similarly to step 1. For each extension $G_{A} = (V, E)$ of a graph $G \in \mathcal{G}_k^{(n - 1)}$ we test that $\text{td}(G_{A}) > k$, as well as testing that $\forall v \in V \colon \text{td}(G_{A} - v) \leq k$.
    In our sequential algorithm we perform the latter test by a simple lookup in $\mathcal{G}_k^{(n - 1)}$.
    The parallel algorithm recomputes the treedepth to save memory at the cost of runtime: Doing so allows us to discard graphs in $\mathcal{G}_k^{(n - 1)}$ as soon as they are no longer needed.

    For \textbf{Step 3}, note that for a graph $G = (V,E) \in \text{Obs}_\subseteq^{(n)}(\mathcal{G}_k) \subseteq \text{Obs}_\sqsubseteq^{(n)}(\mathcal{G}_k)$ it holds that $\forall e \in E \colon G - e \notin \text{Obs}_\sqsubseteq^{(n)}(\mathcal{G}_k)$, since $\text{td}(G - e) \leq k$.
    Conversely, if this property holds for a graph $G \in \text{Obs}_\sqsubseteq^{(n)}(\mathcal{G}_k)$, then any proper subgraph $G' = (V', E')$ of $G$ has treedepth at most $k$:
    If $|V'| < |V|$, then $G'$ is also a subgraph of a proper induced subgraph of $G$, and thus, has treedepth at most $k$.
    If $V' = V, |E'| = |E| - 1$, then each proper induced subgraph of $G'$ is a subgraph of a proper induced subgraph of $G$ and, thus, has treedepth at most $k$.
    Since $G' \notin \text{Obs}_\sqsubseteq^{(n)}(\mathcal{G}_k)$, it follows that $\text{td}(G') \leq k$.
    Finally, if $V' = V, |E'| < |E| - 1$, then $G'$ is a subgraph of $G - e$ for some $e \in E$, which by the previous case already has treedepth at most $k$.

    For \textbf{Step 4}, observe that for a graph $G \in \text{Obs}_\leq^{(n)}(\mathcal{G}_k) \subseteq \text{Obs}_\subseteq^{(n)}(\mathcal{G}_k)$ it holds that $\forall e \in E(G) \colon G / e \notin \text{Obs}_\sqsubseteq^{(n - 1)}(\mathcal{G}_k)$, since $\text{td}(G / e) \leq k$.
    Once more, this property is also sufficient:
    Any minor of $G \in \text{Obs}_\subseteq^{(n)}(\mathcal{G}_k)$ that is also a minor of a proper subgraph of $G$, has treedepth at most $k$.
    Any other minor $G'$ of $G$ is obtained by a series of edge contractions.
    In this case, as in step 3, the condition ensures that $G'$ has treedepth at most $k$.

    \subsection*{Implementation and Results}
    We implemented the algorithm utilizing the nauty library for graph canonization \cite{mckay2014practical} and the Bute library for treedepth computation \cite{trimble:LIPIcs.IPEC.2020.34}.
    Both the source code as well as the computation results are avaiable on Zenodo \cite{kolja_kuehn_2025}.
    We computed $\text{Obs}_\sqsubseteq^{(\leq 16)}(\mathcal{G}_4), \text{Obs}_\subseteq^{(\leq 16)}(\mathcal{G}_4)$ and $\text{Obs}_\leq^{(\leq 16)}(\mathcal{G}_4)$ to be of size 12204, 1718 and 1546 respectively.
    Further computation confirms that $\text{Obs}_\sqsubseteq^{(17)}(\mathcal{G}_4) = \text{Obs}_\sqsubseteq^{(18)}(\mathcal{G}_4) = \emptyset$.
    However, we do not yet have a proof that for all $n > 18$ it holds that $\text{Obs}_\sqsubseteq^{(n)}(\mathcal{G}_4) = \emptyset$.
    
\begin{figure}[h]
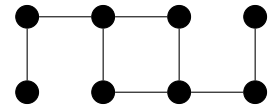

        \centering
        \threeObstruction
        \setlength{\abovecaptionskip}{0pt}
        \setlength{\belowcaptionskip}{-10pt}
        \caption{An induced subgraph obstruction for treedepth 3.}
        \label{fig:missing-obstruction}
    \end{figure}

    In testing the algorithm, we also computed $\text{Obs}_\sqsubseteq(\mathcal{G}_3)$.
    For this, Dvo{\v{r}}{\'a}k et al. \cite{dvovrak2012forbidden} gave a set of 29 graphs.
    However, we found the forbidden induced subgraph shown in Figure \ref{fig:missing-obstruction}, which was overlooked by them.
    As a result,  $|\text{Obs}_\sqsubseteq(\mathcal{G}_3)| = 30$, rather than 29.
    The number of forbidden subgraphs and minors remains at 14 and 12 respectively.
    \setlength{\intextsep}{10pt}

	\bibliographystyle{orcs}
	\bibliography{bibliography.bib}
\end{document}